%% file: main.tex
\documentclass[10pt,conference]{IEEEtran}
\IEEEoverridecommandlockouts
\usepackage{cite}
\usepackage{amsmath,amssymb,amsfonts}
\usepackage{algorithmic}
\usepackage{graphicx}
\usepackage{textcomp}
\usepackage{xcolor}
\def\BibTeX{{\rm B\kern-.05em{\sc i\kern-.025em b}\kern-.08em
    T\kern-.1667em\lower.7ex\hbox{E}\kern-.125emX}}

\usepackage{booktabs}
\usepackage{multirow}
\usepackage{colortbl}
\usepackage{enumitem}
\usepackage{makecell}
\usepackage{subfig}
\usepackage{tcolorbox}
\usepackage[normalem]{ulem}
\usepackage[hidelinks]{hyperref}
    
\begin{document}

\title{An Empirical Study of Architectural Shift from Traditional to AI-Enabled Simulink Controllers\\
\thanks{This work is supported by the National Science Foundation (NSF) under CRII Grant (Award Number: 2347294). The opinions, findings, and conclusions or recommendations presented in this material are those of the author(s) and do not necessarily represent the views of the National Science Foundation (NSF).}
}

\author{\IEEEauthorblockN{Hadiza Umar Yusuf}
\IEEEauthorblockA{%\textit{Dept. of Computer and Information Science} \\
\textit{University of Michigan-Dearborn}\\
Dearborn, USA. \\
hyusuf@umich.edu}
\and
\IEEEauthorblockN{Khouloud Gaaloul}
\IEEEauthorblockA{%\textit{Dept. of Computer and Information Science} \\
\textit{University of Michigan-Dearborn}\\
Dearborn, USA.\\
kgaaloul@umich.edu}
% \and
% \and
% \and
}

\maketitle

\begin{abstract}
Effective AI adoption in cyber-physical systems (CPS) depends on embedding design knowledge into engineering practice. Yet as AI-enabled components increasingly replace analytically derived control laws, this occurs without a systematic understanding of how controller architectures differ or remain similar across paradigms. We address this gap with an empirical study of traditional and AI-enabled Simulink controllers, guided by a literature-derived taxonomy of ten structural categories and nine functional roles. The study analyzes 62 real-world models spanning 8 controller types and 10 application domains, and surveys 13 practitioners, identifying three architectural tensions. First, subsystem organization dominates all controller structures regardless of paradigm, occupying 68--72\% of controller footprint, while core control logic occupies minimal space. Second, AI-enabled controllers rely heavily on discrete dynamics and user-defined abstraction, categories largely absent from AI literature, exposing a gap between described and implemented architectures. Third, constraint enforcement blocks largely disappear from AI-enabled models despite practitioner expectations. This reveals a misalignment where safety mechanisms shift from explicit structure to implicit training-time artifacts, breaking traceability.
\end{abstract}

\begin{IEEEkeywords}
AI-enabled systems, controller architecture, cyber-physical systems, deep reinforcement learning, empirical study, model-based design, simulink, taxonomy.
\end{IEEEkeywords}

\section{Introduction}
\label{sec:introduction}
\input{Sections/Introduction}

\section{Background}
\label{sec:background}
\input{Sections/Background}

\section{Literature-Derived Taxonomy Construction}
\label{sec:approach}
\input{Sections/Approach}

\section{Empirical Evaluation}
\label{sec:setup}
\input{Sections/Setup}

%\section{Evaluation results}
%\label{sec:evaluation}
\input{Sections/Evaluation}

\section{Threats to Validity}
\label{sec:treats}
\input{Sections/Threats}

\section{Related Work}
\label{sec:related}
\input{Sections/Related}

\section{Conclusion}
\label{sec:conclusion}
\input{Sections/Conclusion}

\bibliographystyle{IEEEtran}
\bibliography{bibliography}

\end{document}

%% file: Sections/Introduction.tex
The integration of Artificial Intelligence (AI) into system development has given rise to new research areas. These include AI for Model-Based Systems Engineering (MBSE)~\cite{Song2025AI4MBSE} and Model-Driven Engineering (MDE) for AI~\cite{Radler2024MDEAI}. Recent work argues that the real impact requires redesigning engineering workflows and incorporating architectural awareness, not just better tools~\cite{10176194}. This is because effective AI adoption depends on embedding design knowledge into engineering practice. This is particularly important for Cyber-Physical Systems (CPS), which integrate computation, control, and physical processes to enable intelligent, autonomous behavior across domains~\cite{baheti2011cyber}. Simulink has become the dominant modeling environment for CPS development in automotive, aerospace, and industrial applications~\cite{Boll2021OpenSourceSimulink}. Simulink models are central artifacts for controller design, analysis, testing, and deployment. They often provide the most direct representation of how control logic is structurally realized in practice. Simulink-based CPS development has been extensively studied from testing, analysis, and variability perspectives~\cite{Elberzhager2013SimulinkMapping, Weiland2014SimulinkVariabilityClassification}. Far less attention has been given to controller architecture itself, especially how AI-enabled controllers differ structurally from traditional designs~\cite{hadiza2025Nav}.

In traditional CPS engineering, controllers are typically derived from explicit mathematical models of system dynamics. Control laws are designed analytically using approaches such as Model Predictive Control (MPC), Proportional-Integral-Derivative (PID) control, and Linear Quadratic Regulators (LQR)~\cite{Okasha2022DesignAE,Varma2020TrajectoryTO}. More recently, CPS control design has shifted toward AI-based approaches, particularly Deep Reinforcement Learning and Neural Networks~\cite{lee2022deep,li2023deep}. In these approaches, control strategies are learned from data or interaction, and learning-based components increasingly serve as core decision-making and control elements~\cite{bucsoniu2018reinforcement, lee2022deep}. This shift offers greater flexibility but changes how control logic is represented in engineering models. As noted in a recent study, AI-enabled systems ``behave like conventional systems, right up until they don't''~\cite{ozkaya2025preliminary}. This highlights the need for architectural characterization of both paradigms and for understanding the tensions between literature and engineering practice.

Despite the growing use of AI-enabled control in CPS, the literature still lacks a shared architectural lens for comparing traditional and AI-enabled controllers. This gap is notable because Simulink is the environment where these controllers are actually designed and verified. Existing comparisons largely focus on behavioral outcomes, such as performance, robustness, or safety violations~\cite{song2022cyber,lee2022deep,Okafor2021RLBasedPIDvsClassical}, using metrics like tracking error, stability margins, or violation rates. These comparisons treat controller architecture as an implicit backdrop, not an explicit object of analysis. To the best of our knowledge, no prior work has described the systematic structural differences between traditional and AI-enabled controllers. This leaves a gap in the literature, making it difficult to reason about design complexity across control paradigms. 

This paper addresses this gap through an empirical study of traditional and AI-enabled controller architectures in Simulink-based CPS. The study is guided by a literature-derived taxonomy of ten structural categories (C1--C10) and nine functional roles (L1--L9), providing a consistent vocabulary independent of paradigm. Using this taxonomy, we study the shift from traditional to AI-enabled design in both literature and practice. We analyze 62 real-world Simulink models spanning 8 controller types and 10 application domains, supported by a practitioner survey. This is the first empirical, cross-paradigm characterization of controller architecture, drawn from literature, real-world systems, and practitioner perspectives. It reveals three tensions that establish the first empirical baseline for cross-paradigm evaluation. First, structural scaffolding, particularly subsystem organization (C6), dominates all controllers regardless of paradigm. Second, AI-enabled controllers rely heavily on discrete dynamics (C2) and user-defined abstraction (C8) despite these being nearly absent in AI literature, exposing a documentation gap. Third, constraint enforcement blocks (C3) largely disappear from AI-enabled models even as practitioners expect them, revealing a misalignment between training-time safety and deploy-time structure. Our contributions are summarized as follows:

\begin{itemize}
    \item We construct a literature-derived taxonomy of structural categories and functional roles that serves as a unified instrument for cross-paradigm architectural comparison.

    \item We characterize and compare controller composition from two complementary perspectives: how traditional and AI-enabled controllers are portrayed in the literature, and how they are structurally realized in 62 real-world Simulink models, revealing systematic differences between documented and practiced architecture.

    \item We corroborate these patterns through a survey-based manual assessment involving 13 practitioners, examining whether the identified categories align with practitioner judgment and discussing implications for software engineering practice.
    
    \item We identify three architectural tensions across literature, system models, and practitioner perception, and discuss their implications for MBSE tool support.
\end{itemize}

\textbf{Structure.} Section~\ref{sec:background} introduces AI-enabled CPS as this paper's primary context. Section~\ref{sec:approach} describes the construction of our literature-derived taxonomy of controller architectures in Simulink-based CPS. Section~\ref{sec:setup} formalizes the research questions, describes our dataset, model characterization, and survey design, and analyzes the evaluation results. Section~\ref{sec:treats} discusses threats to validity, Section~\ref{sec:related} reviews related work, and Section~\ref{sec:conclusion} concludes the paper.

%These controllers expose dynamics, constraints, and supervisory logic directly in their structural realization, which supports interpretability, stability analysis, and verification. However, their reliance on accurate modeling assumptions and manual tuning can limit adaptability in complex or only partially understood environments~\cite{chatzilygeroudis2020benchmark}.

%% file: Sections/Background.tex
\begin{figure}[t]
\centering
\includegraphics[width=0.5\textwidth]{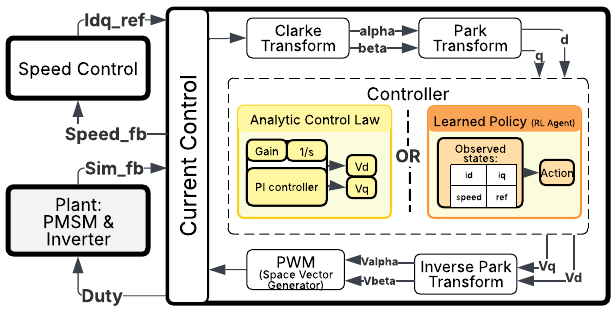}
\caption{Illustrative PMSM field-oriented control (FOC) example: a learned policy block replaces the analytic control law while the surrounding structure stays unchanged.}
\label{fig:pmsm}
\end{figure}

% Simulation-Based Design
Model-Based Design (MBD) has long been the foundation of CPS development, enabling early-stage modeling, simulation, and verification prior to deployment~\cite{nicolescu2018model}. The Simulink/Stateflow toolset provides domain-specific libraries for designing, analyzing, and validating control-intensive systems across automotive, aerospace, and energy domains~\cite{MathWorks2023,mavridou2020ten}. Controller logic is built from interconnected atomic blocks organized hierarchically into subsystems. Simulink itself is the modeling environment in which controllers are implemented. The object of our study is the block-level structure of Simulink models, not the tool or platform itself.

% Controller Paradigms
We refer to controllers historically derived from analytical paradigms as \emph{traditional controllers}. These include classical linear and optimal techniques such as PID, LQR, and MPC. They also include supervisory or rule-based controllers such as Conditional Logic Controllers (CLC) and State-Based Controllers (SBC)~\cite{Okasha2022DesignAE,Varma2020TrajectoryTO,ramadge1987supervisory}. All rely on explicit mathematical models and predefined control laws. In contrast, \emph{AI-enabled controllers} incorporate data-driven components, such as Deep Reinforcement Learning (DRL), Deep Neural Networks (DNN), or Fuzzy Logic Controllers (FLC)~\cite{bucsoniu2018reinforcement}. These approximate control policies from interaction or data, encapsulating behavior within learned components instead of analytically encoded logic.

% Why Controller Structure Matters
AI's growing integration within CPS introduces architectural shifts that are not merely algorithmic but structural~\cite{song2022cyber,xie2023mosaic,hadiza2025Nav}, affecting maintainability, interpretability, assurance, and reuse. This is because safety and robustness assessments in model-based CPS development depend on structural decomposition and traceability across subsystems~\cite{nejati2019evaluating}. AI-enabled controllers often encapsulate decision logic within learned policy blocks or neural network components, reducing explicit traceability and complicating safety analysis~\cite{huang2019reachnn,tran2020nnv}. This is an increasingly important concern as CPS operate in safety-critical domains governed by standards such as ISO~21448 (SOTIF)~\cite{no202221448}.
% Structural composition influences verification complexity, test coverage strategies, fault localization, and integration with formal analysis tools.

To illustrate the architectural contrast between paradigms, consider the field-oriented control of a permanent magnet synchronous motor (PMSM) implemented in Simulink (Figure~\ref{fig:pmsm}). In the traditional implementation, the current-control subsystem uses deterministic control-law blocks encoding system dynamics and error correction, with saturation blocks enforcing current limits. The AI-enabled implementation instead replaces this analytic structure with a learned policy block, while the surrounding architecture remains unchanged. This highlights the central issue: replacing a traditional controller with an AI-enabled one changes not only the control algorithm. It also changes the internal structural organization through which control responsibilities are realized, including how constraints are enforced. This challenge is amplified by the diversity of controller types within both paradigms, each realizing control intent through different model constructs. This motivates a systematic taxonomy-guided characterization for comparing controller architectures across paradigms.

%% file: Sections/Approach.tex
\begin{figure}[t]
    \centering
    \includegraphics[width=\columnwidth]{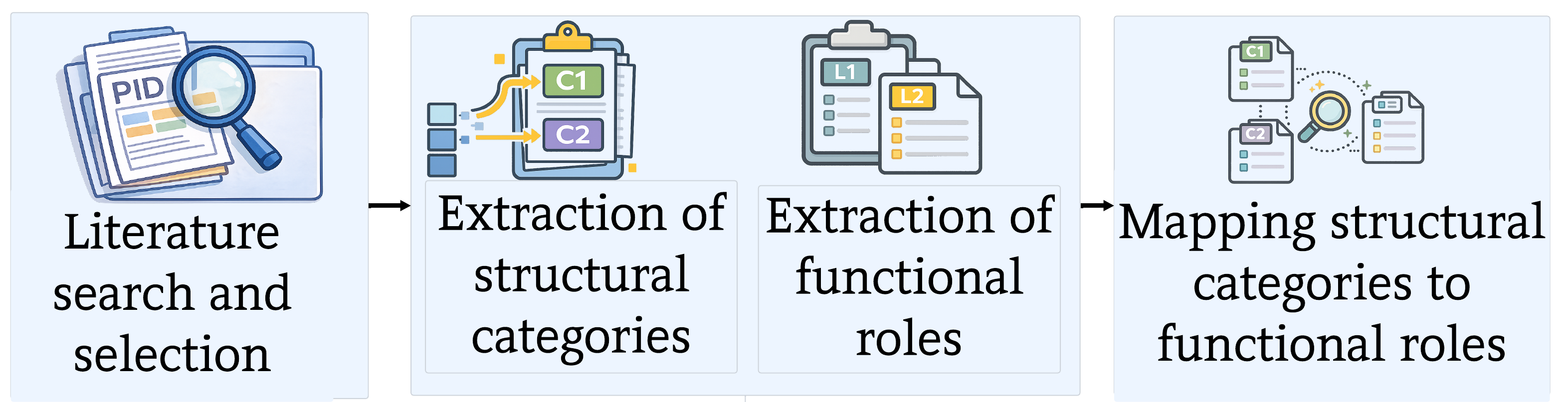}
    \caption{Taxonomy construction process} 
    \label{fig:methodology}
\end{figure}

Building on the architectural shift illustrated earlier, we conduct a systematic empirical characterization guided by a literature-derived taxonomy. This taxonomy is a structured classification of Simulink block types, organized by their structural and functional role in controller design. Figure~\ref{fig:methodology} illustrates the taxonomy construction process, which we detail through the following steps.

\textit{Step 1: Literature Search and Selection.}
This step curates a corpus of literature discussing controller design in the context of MATLAB/Simulink. We apply three inclusion criteria for a source to be included:
\begin{enumerate}
    \item it must explicitly describe controller structure, block-level implementation, or Simulink modeling guidance;
    \item it must be authoritative, that is, a peer-reviewed publication, official MathWorks documentation, or a widely adopted textbook; and
    \item it must be specific to a control paradigm (traditional or AI-enabled) or to paradigm-agnostic Simulink conventions.
\end{enumerate}
The search was conducted manually across multiple academic and technical databases, guided by these criteria. We exclude informal grey literature such as blog posts, forum discussions, and unreviewed tutorials. Our grey literature is restricted to Tier 1 sources under Garousi et al.'s multivocal literature taxonomy~\cite{garousi2019guidelines}: official vendor documentation and modeling guideline reports. We then search for canonical references covering established controller types in both paradigms. These include PID, LQR, MPC, and state-space control for traditional controllers, and DRL, DNN, and FLC for AI-enabled controllers. We also include general Simulink and model-based design guideline sources, namely the Simulink User's Guide and the MathWorks Automotive Advisory Board (MAB) Modeling Guidelines. These guideline sources ground paradigm-agnostic categories such as subsystem organization, signal routing, and interfacing. The resulting corpus comprises 21 sources (4 traditional-control, 13 AI-control, 4 general guideline), with the complete list provided in our replication package~\cite{anonymous_zenodo_2026}.

This split is uneven because of the underlying literature's composition, not sampling bias: a single textbook chapter can span PID, LQR, and state-space design, while AI-based control spans more distinct techniques, RL, several DRL variants, DNNs, and fuzzy logic, documented independently across dedicated papers and toolboxes.

% Sources are selected to reflect authoritative guidance and widely cited practice, including the MathWorks Automotive Advisory Board (MAB) Guidelines, Modeling Guidelines for Simulink, the Reinforcement Learning Toolbox User Guide, and foundational textbooks and survey papers on classical, supervisory, and AI-based control. The search focuses on sources that describe controller structure, modeling guidelines, or architectural organization. The resulting corpus comprises 17 sources.

\begin{table*}[!t]
    \centering
    \caption{Simulink Block Categories and Associated Block Types~\cite{MathWorks2024}. The complete block types are available at~\cite{anonymous_zenodo_2026}.}
    \label{tab:matlablibrary}
    \fontsize{8}{10}\selectfont
    \setlength{\tabcolsep}{2.8pt}
    \centering
    % \begin{tabular}{@{}p{0.03\textwidth} p{0.23\textwidth} p{0.71\textwidth}@{}}
    % {p{0.02\linewidth}p{0.22\linewidth}p{0.68\linewidth}}
    \begin{tabular}{l l l}
    \toprule
    \textbf{ID} & \textbf{Categories} & \textbf{Block Types} \\
    \midrule
    C1 & Continuous Dynamics & Derivative, Integrator,  PID Controller, State-Space,  Transfer Fcn, Transport Delay, Zero-Pole, FOH, etc.\\
    % \hline
    C2 & Discrete Dynamics \& State & Delay, Difference, Discrete Derivative, Discrete PID Controller, Discrete State-Space, Memory, ZOH, etc.\\
    % \hline
    C3 & Discontinuities \& Nonlinearities & Coulomb and Viscous Friction, Dead Zone, Hit Crossing, Rate Limiter, Relay, Saturation, Backlash, etc.\\
    % \hline
    C4 & Logic, Conditions, \& Events & Combinatorial Logic, Compare To Zero, Logical and Relational Operator, Shift Arithmetic, Extract Bits, etc.\\
    % \hline
    C5 & Math \& Signal Operations & Abs, Algebraic Constraint, Sum, Bias, Divide, Dot Product, Gain, Math Function, Reshape, Sign, etc. \\
    % \hline
    C6 & Ports, Interfaces \& Subsystems & Enable, Subsystem, If, If Action Subsystem, Inport, Outport, Switch Case, Trigger, In/Out Bus Element, etc. \\
    % \hline
    C7 & Signal Sources \& Stimuli & Clock, Constant, From File/Workspace, Ground, Pulse Generator, Ramp, Random Number, Sine Wave, etc.	\\
    % \hline
    C8 & User-Defined Logic \& Abstraction & C Function, Fcn, MATLAB Function, MATLAB System, Reset Function, S-Function, S-Function Builder, etc.\\
    % \hline
    C9 & Memory \& Data Buffers & Bus Assignment/Creator/selector, Data Store Memory/Read/Write, Mux, Demux, From/Goto, Merge, etc.\\
    % \hline
    C10 & Artificial Intelligence & RL Agent, Policy, Predict, Image Classifier, Deep Neural Networks, ONNX Predict, Fuzzy Logic, etc.\\
    \hline
    \end{tabular}
\end{table*}

\textit{Step 2: Literature Extraction of Structural Categories and Functional Roles.}
A category groups Simulink block types serving a similar modeling purpose in controller implementation, formed by organizing controller-relevant blocks by implementation domain and modeling intent. Blocks explicitly referenced in the literature are mapped to Simulink's native library organization~\cite{MathWorks2024}. They are then grouped into categories capturing continuous and discrete dynamics, signal computation, logic and event handling, data routing, and AI-specific computation. For each candidate block, we distinguish explicit references, where the source names the block directly, from inferred references. Inferred references describe a function or mechanism the block canonically implements without naming it. Our categorization follows an iterative process consistent with established taxonomy development methods~\cite{nickerson2013method}. This process alternates between two directions. A conceptual-to-empirical direction proposes categories from controller design principles in the literature. An empirical-to-conceptual direction then checks them against Simulink's library structure. This confirms that categories correspond to actual block groupings, not abstract distinctions with no implementation counterpart. We stop once two conditions are met. First, no additional category is needed for a newly examined block or source (an objective condition). Second, the categories are mutually exclusive and collectively exhaustive over our corpus (a subjective condition). The process resulted in ten structural categories (C1--C10), summarized in Table~\ref{tab:matlablibrary}.

% The categorization follows a bidirectional process in which candidate categories are aligned with Simulink’s library structure and validated against literature-referenced blocks to ensure complete coverage.

 % (\href{https://www.mathworks.com/help/simulink/referencelist.html?type=block&category=block-libraries&s_tid=CRUX_topnav}{Link to reference})

\begin{table*}[!t]
\centering
\caption{Functional Role Levels (L1–L9) Derived for Controller Characterization. Each level describes a distinct functional purpose within Simulink-based CPS architectures.}
\label{tab:functionalroles}
\fontsize{8}{10}\selectfont
\setlength{\tabcolsep}{2.5pt}
\begin{tabular}{l l l}
% {@{}p{0.010\textwidth} p{0.23\textwidth} p{0.71\textwidth}@{}}
\toprule
\textbf{ID} & \textbf{Level Type} & \textbf{Description} \\
\midrule
L1 & Core Control Dynamics & Implements the primary control law that generates control actions from system inputs and states. \\
% \hline
L2 & Signal Constraint Enforcement & Applies constraints or limits to control signals to ensure safe, stable, and valid operation of the system. \\
% \hline
L3 & Event/Mode Switching Logic & Enables discrete changes in control behavior based on events, thresholds, or operating conditions. \\
% \hline
L4 & Signal Preprocessing \& Computation & Performs transformations or computations to ensure signal compatibility within the control architecture. \\
% \hline
L5 & Supervisory Coordination & Manages coordination between multiple control elements or modes, ensuring proper sequencing and integration. \\
% \hline
L6 & Estimation / Observer Support & Provides state estimation or variable reconstruction to supply information not directly measured. \\
% \hline
L7 & Optimization / Reference Shaping & Generates optimized trajectories, setpoints, or reference signals for the control system to follow. \\
% \hline
L8 & Learning \& Adaptation & Supports modification, tuning, or training of control strategies based on data or evolving system behavior. \\
% \hline
L9 & Environment \& External Interaction & Handles communication and interaction with external systems, physical plants, or simulation environments. \\
\bottomrule
\end{tabular}
\end{table*}

Functional roles describe what a component does in the control pipeline, independent of the specific Simulink block types used. We derive functional roles by extracting recurring descriptions of controller responsibilities and architectural intent from the literature, then abstracting them into role definitions. For example, classical control texts describe controllers as mappings from system states or error signals to control actions. This is commonly expressed as $u(t)= -Kx(t)$ for state-feedback control or $u(t)=f(e(t))$ for error-based control. These mappings correspond to Simulink realizations built from state-space, gain, integrator, and PID structures \cite{Tewari2002}. This step produced nine functional role levels (L1--L9), summarized in Table~\ref{tab:functionalroles}.

\textit{Step 3: Mapping Structural Categories to Functional Roles.}
We map structural categories to functional roles using block-level literature descriptions. For each explicitly referenced block, we assign one or more functional roles based on how the literature characterizes its canonical use. This captures how categories support specific control responsibilities. Unlike categories, which are assigned exclusively per block, functional roles may overlap. A block occupies one structural position but can serve multiple functional purposes in the control pipeline. This distinction reflects a structural reality: a block occupies one position by construction, but its behavioral contribution to control can serve multiple purposes. For example, consider the Saturation block, which appears explicitly in traditional control literature as enforcing current or signal limits. It is assigned to Discontinuities and Nonlinearities (C3), the structural category covering constraint-related blocks. Because Saturation applies limits to control signals to ensure safe operation, it is also assigned the functional role Signal Constraint Enforcement (L2). Category and role assignments, with justification and source page reference, are recorded per block in our replication package~\cite{anonymous_zenodo_2026}.

%% file: Sections/Setup.tex
\subsection{Research Questions}
\label{sec:rqs}

Our empirical evaluation examines how controller composition differs between traditional and AI-enabled paradigms, in literature and in practice. We structure this around three research questions: the first two drive the central comparison, while the third corroborates the results through practitioner assessment.
\begin{itemize} %[topsep=2pt,itemsep=2pt,parsep=0pt,leftmargin=*]
    \item RQ1 (Literature-based composition). \textit{How does the literature's portrayal of controller structure shift between traditional and AI-enabled paradigms? Specifically, we ask which categories and functional roles receive emphasis in each, and what structural elements are present in one paradigm's literature but absent in the other.} This question captures both emphasis and omission in how each paradigm's literature describes controller design.
    \item RQ2 (Model-based composition) \textit{What taxonomy categories are most prevalent and structurally dominant in real Simulink controller subsystems, and how do these empirical patterns align with or diverge from the literature-derived expectations from RQ1?} This question examines what engineers actually use in practice, compared against RQ1's literature-derived expectations. We treat model-based composition and its comparison to RQ1 as a single question, since both rely on the same underlying measurements of category presence.
    \item RQ3 (Practice-based corroboration). \textit{To what extent does manual assessment by software engineering practitioners corroborate the structural differences between traditional and AI-enabled controllers identified in RQ1 and RQ2, and what implications do these differences carry for software engineering practice?} This question checks whether the composition patterns identified through RQ1 and RQ2 are also recognizable to practitioners, and what those patterns mean for practice.
\end{itemize}

% investigates if practitioners are able to identify which categories are essential and dominant in each paradigm, and assesses their judgment against the literature-based and the model-based composition. 

\subsection{Empirical Evaluation Plan}
\label{sec:exp_setup}

To answer the research questions, we use the taxonomy synthesized in Section~\ref{sec:approach} in three stages. First, we analyze the literature-based composition of traditional and AI-enabled controller architectures as explicitly described (RQ1). Second, we quantitatively assess model-based composition of real-world systems drawn from diverse CPS domains (RQ2). Finally, we corroborate the resulting patterns through practitioner-based manual assessment (RQ3). 

% Overall, we examine 62 Simulink CPS models with respect to the taxonomy categories and functional roles, allowing us to assess alignment between real-world design practices and literature patterns, and to derive insights into existing gaps and opportunities for improving CPS model design and analysis.

% \begin{figure}[t]
%     \centering
%     \subfloat[Distribution of controller types]{
%         \includegraphics[width=0.24\textwidth]{images/model_stats01_.png}
%         \label{fig:modelstats_controller}
%     }%\hfill
%     \subfloat[Distribution of domains]{
%         \includegraphics[width=0.25\textwidth]{images/model_stats02_.png}
%         \label{fig:modelstats_domain}
%     }
%     \caption{Model statistics for the CPS benchmark: distribution of controller types (AI-based vs.\ traditional) and application domains.}
%     \label{fig:modelstats_summary}
% \end{figure}

\textbf{Dataset Curation.} To answer RQ2, we curated a dataset of Simulink-based CPS models from open-source repositories. Sources include prior work in CPS, artificial intelligence, and software engineering research~\cite{zhang2020hybrid,zhang2022falsifai,song2022cyber,lyu2023autorepair,nejati2019evaluating,huang2019reachnn,tran2020nnv,althoff2015introduction,xiang2018output,song2023mathtt,xie2023mosaic}, MathWorks example repositories, and CPS verification competition benchmarks~\cite{ernst2022arch,ernst2021arch,johnson2021arch}. We screened all models for relevance, completeness, and the presence of an explicitly identifiable controller subsystem. Each model was classified as AI-enabled if its control law was realized through a learned component, such as a DRL agent, DNN, or FLC block. Otherwise, it was classified as traditional. The final dataset contains 62 Simulink models spanning multiple CPS domains. It is balanced across 31 traditional controllers (11 PID, 6 MPC, 4 LQR, 3 SBC, 7 CLC). The remaining 31 are AI-enabled controllers (23 DRL, 3 DNN, 5 FLC). The complete list is provided in the replication package~\cite{anonymous_zenodo_2026}. For each system model, we isolate blocks belonging to the controller subsystem, excluding blocks associated with the plant, environment, or auxiliary simulation components. We then perform exhaustive block extraction traversing nested subsystems, masked components, linked library elements, and all variant configurations~\cite{MathWorksfindsystem}, ensuring full controller capture beyond top-level visibility. Extracted blocks are then assigned to taxonomy categories according to our established definitions. 

\textbf{Model Characterization.} We apply the literature-derived taxonomy to the extracted controller subsystems by detecting the presence of taxonomy categories (C1 to C10) based on the Simulink block types they contain, without interpreting execution semantics or control behavior. Category presence is recorded uniformly at the controller level, based on whether a category appears at least once in the subsystem. This abstraction supports fair cross-model comparison by reducing sensitivity to differences in controller complexity, modeling granularity, and domain-specific implementation style.

\textbf{Metrics.} Our quantitative analysis uses block-level category assignments across three complementary metrics, capturing structural space, within-model prominence, and cross-model adoption. Let $N$ denote the number of models, $B_i$ the number of controller blocks in model $i$, and $b_{i,c}$ the number of blocks in category $c$.

\emph{Coverage} $\mathit{Cov_c}$ measures dataset-level structural emphasis on each category, useful for comparing literature-reported and real-world patterns since it reflects aggregate design emphasis, not per-model variation:
\begin{equation}
\mathit{Cov_c} = \frac{\sum_{i=1}^{N} b_{i,c}}{\sum_{i=1}^{N} B_i} \times 100\%.
\end{equation}

\emph{Mean Normalized Presence} $\mathit{MNP_c}$ measures average within-model prominence by computing each category's relative share per model, then averaging across models:
\begin{equation}
\mathit{MNP_c} = \frac{1}{N} \sum_{i=1}^{N} \frac{b_{i,c}}{B_i}.
\end{equation}
$\mathit{MNP_c}$ determines how central a category is within individual controllers, weighting each model equally regardless of size and better reflecting typical structural importance than Coverage alone.

\emph{Model-Level Prevalence} $\mathit{MLP_c}$ measures category adoption as the proportion of models containing at least one instance:
\begin{equation}
\mathit{MLP_c} = \frac{1}{N} \sum_{i=1}^{N} I(b_{i,c} > 0) \times 100\%,
\end{equation}
where $I(\cdot)$ equals 1 if category $c$ appears in model $i$ and 0 otherwise. $\mathit{MLP_c}$ captures how broadly a category is adopted across the dataset, distinguishing common ingredients from specialized ones regardless of block-count contribution.

\begin{figure}[t]
    \centering
    \includegraphics[width=0.5\textwidth]{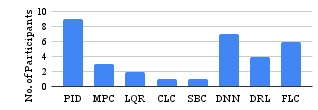}
    \caption{Participants' experience of controller designs} 
    \label{fig:esp}
\end{figure}

\begin{table}[t]
\centering
\caption{12 representative Simulink CPS models used in the survey. Size denotes controller subsystem block count.}
\label{tab:models}
% \small
\scriptsize
\renewcommand{\arraystretch}{1.4}
\setlength{\tabcolsep}{1.4pt}
% \begin{tabular}{@{}p{0.010\textwidth} p{0.15\textwidth} p{0.010\textwidth} p{0.010\textwidth}|p{0.010\textwidth} p{0.15\textwidth} p{0.010\textwidth} p{0.010\textwidth}@{}}
\begin{tabular}{l l l r | l l l r}
\toprule
\textbf{ID} & \textbf{Model Name} & \textbf{Type} & \textbf{Size} & \textbf{ID} & \textbf{Model Name} & \textbf{Type} & \textbf{Size}\\
\midrule
M01 & Abstract Fuel Control & DRL & 400 
& M07 & PMSM Control & PID & 2601 \\
M02 & Neural Network & DNN & 699
& M08 & Stochastic Fault Tolerance & SBC & 273 \\
M03 & Temperature Control & FLC & 107 
& M09 & Adaptive Cruise Control & MPC & 361 \\
M04 & Artificial Pancreas Control & FLC & 276 
& M10 & Missile Guidance System & PID & 396 \\
M05 & Steam Condenser & DNN & 172 
& M11 & Rolling Mill & LQR & 211 \\
M06 & Nonlinear Guidance & CLC & 355 
& M12 & House Heating System & MPC & 200 \\
\bottomrule
\end{tabular}
\end{table}

\textbf{Survey.} To corroborate the literature-based and model-based composition patterns with human judgment, we conduct an anonymous survey. This survey examines how individuals interpret controller structure when presented with Simulink-based system models and block category groupings. We recruit 13 participants through colleagues and professional networks with controller-design experience reported in Figure~\ref{fig:esp}. Surveying all 62 models was impractical, so we selected 12 representative models based on internal design style (Table~\ref{tab:models}). These models cover three controller representation styles:
\begin{itemize}
    \item encapsulated designs using existing Simulink masked blocks;
    \item equation-based designs realized through explicit dynamic and mathematical blocks; and
    \item function-based designs using MATLAB Function or S-Function blocks.
\end{itemize}
The selected models span all controller types and both paradigms, reflecting the full dataset's structural diversity.

The survey has three parts. The first introduces the study goals and collects participant background information, including self-reported experience with Simulink-based system modeling (Figure~\ref{fig:esp}). The second presents the main assessment per model: a dominance rating (0--2 scale), an essentiality rating (1--5 scale), a paradigm classification, and an open-ended justification. The third collects post-study reflections on the block group framework's clarity and usability. The survey took approximately 60 minutes per participant.

To support consistent interpretation, taxonomy categories are presented as five labeled block groups. C5 and C6 are excluded, as both appear ubiquitously regardless of paradigm. C2 and C4 are merged, governing discrete-time and event-driven behavior. C7 and C9 are merged, managing data availability within the controller, and C1, C3, and C8 are presented individually. We define \emph{architectural scaffolding} as structural infrastructure that integrates the control component into the broader system without implementing control logic. C5 and C6 fall into this category. Full group definitions appear in our replication package~\cite{anonymous_zenodo_2026}. Participants see Simulink screenshots of each model's controller view with unlabeled groupings.

RQ1 is addressed by computing $\mathit{Cov_c}$ to quantify structural emphasis on taxonomy categories in literature-based descriptions. RQ2 is addressed by computing $\mathit{Cov_c}$, $\mathit{MNP_c}$, and $\mathit{MLP_c}$ over the $62$ models to characterize real-world structural composition. RQ3 is addressed by analyzing survey responses for practitioner corroboration of these patterns and their software engineering practice implications.

\textbf{Data Availability.} All data and artifacts are available in our replication package~\cite{anonymous_zenodo_2026}.

%Block extraction is performed using MATLAB’sfunction~\cite{MathWorksfindsystem}, configured with \emph{LookUnderMasks} set to \emph{'all'}, \emph{FollowLinks} set to \emph{'on'}, and \emph{MatchFilter} set to \emph{Simulink.match.allVariants}. These settings ensure complete enumeration of controller blocks, including those embedded within subsystems, masked components, and variant configurations. Extracted blocks are mapped to taxonomy categories using the definitions established in the approach.

%% file: Sections/Evaluation.tex
\begin{table}[!t]
\centering
\caption{Literature-derived block counts (BC) and coverage $\mathit{Cov_c}$ for taxonomy categories (C1–C10) and functional-role levels (L1–L9) in traditional (T) and AI-enabled (AI) controller descriptions.}
\label{tab:rq1_lit_compare_roles}
\scriptsize
% \small
\setlength{\tabcolsep}{1.5pt}
\renewcommand{\arraystretch}{1.2}
\begin{tabular}{c c r r r| c r r r r}
\toprule
\textbf{Cat} 
& \textbf{BC(T)} 
& \textbf{$\mathit{Cov_c}$(T)} 
& \textbf{BC(AI)} 
& \textbf{$\mathit{Cov_c}$(AI)} 
& \textbf{FRL} 
& \textbf{BC(T)} 
& \textbf{$\mathit{Cov_c}$(T)} 
& \textbf{BC(AI)} 
& \textbf{$\mathit{Cov_c}$(AI)}\\
\midrule
C1 & \cellcolor{yellow}8 & \cellcolor{yellow}25.8\% & 0 & 0.0\% &
L1 & \cellcolor{yellow}12 & \cellcolor{yellow}38.7\% & \cellcolor{blue!30}10 & \cellcolor{blue!30}35.7\% \\

C2 & \cellcolor{yellow}7 & \cellcolor{yellow}22.6\% & \cellcolor{blue!30}5 & \cellcolor{blue!30}17.9\% &
L2 & \cellcolor{yellow}5 & \cellcolor{yellow}16.1\% & 0 & 0.0\% \\

C3  & \cellcolor{yellow}6 & \cellcolor{yellow}19.4\% & 0 & 0.0\% &
L3 & 2 & 6.5\% & 2 & 7.1\% \\

C4  & 0 & 0.0\% & 1 & 3.6\% & 
L4 & \cellcolor{yellow}5 & \cellcolor{yellow}16.1\% & \cellcolor{blue!30}8 & \cellcolor{blue!30}28.6\% \\

C5 & 2 & 6.5\% & 1 & 3.6\% & 
L5 & 1 & 3.2\% & 2 & 7.1\% \\

C6  & 3 & 9.7\% & 3 & 10.7\% & 
L6 & 0 & 0.0\% & 0 & 0.0\% \\

C7  & 2 & 6.5\% & 2 & 7.1\% & 
L7 & 2 & 6.5\% & 1 & 3.6\% \\

C8  & 2 & 6.5\% & \cellcolor{blue!30}3 & \cellcolor{blue!30}10.7\% & 
L8 & 2 & 6.5\% & \cellcolor{blue!30}4 & \cellcolor{blue!30}14.3\% \\

C9  & 1 & 3.2\% & \cellcolor{blue!30}4 & \cellcolor{blue!30}14.3\% & 
L9 & 2 & 6.5\% & 3 & 10.7\% \\

C10  & 0 & 0.0\% & \cellcolor{blue!30}9 & \cellcolor{blue!30}32.1\% & 
   &  &  &  &\\
\midrule
\textbf{Total} 
& \textbf{31} 
& \textbf{100.0\%} 
& \textbf{28} 
& \textbf{100.0\%} 
& \textbf{Total} 
& \textbf{31} & -- 
& \textbf{28} & -- \\
\bottomrule
\end{tabular}
\footnotesize{\\
\textit{Note:} Functional-role coverage for AI exceed 100\% in aggregate because some blocks are associated with multiple roles. }
\end{table}

\subsection{Results}
\subsubsection{RQ1: Literature-based composition}
\label{sec:rq1}

Table~\ref{tab:rq1_lit_compare_roles} summarizes the structural emphasis reported in the literature for traditional and AI-enabled controllers, using the proposed taxonomy. It reports the absolute counts $BC(T)$ and $BC(AI)$. These denote the number of distinct block types explicitly or inferentially referenced in traditional and AI-enabled controller literature, respectively. This follows the explicit and inferred distinction established in Section~\ref{sec:approach}. It also reports $\mathit{Cov_c}(T)$ and $\mathit{Cov_c}(AI)$ for each category and functional-role level.

Traditional control literature is dynamics centered, concentrating strongly on dynamic and constraint-related structure. Continuous dynamics (C1, $25.8\%$), discrete dynamics and state (C2, $22.6\%$), and discontinuities and nonlinearities (C3, $19.4\%$) are the most referenced categories. These three account for about $2/3$ of all explicitly referenced traditional blocks. At the block level, this emphasis appears in references to \emph{Transfer Function}, \emph{State-space}, \emph{Integrator}, and \emph{Derivative} blocks of C1 and C2. These blocks encode system dynamics and state evolution. It also appears in \emph{Saturation}, \emph{Rate Limiter}, and \emph{Dead Zone} blocks of C3, used for constraint handling. These blocks form the dominant elements through which traditional control literature specifies controller behavior, reflecting a design focused on mathematical dynamics and constraint.

AI-enabled control literature is computation and integration-centered, placing dominant emphasis on AI-specific constructs (C10, $32.1\%$), the single largest category in reference. It also shows notable presence of user-defined logic (C8, $10.7\%$) and memory/data structures (C9, $14.3\%$). This focus appears in references to composite AI components such as \emph{RL Agent}, \emph{Neural Network Predict}, and \emph{Fuzzy Logic Controller} blocks (C10). These blocks encapsulate learned control behavior. It also appears in supporting structures such as \emph{MATLAB Function} blocks (C8) and data-handling elements including \emph{Mux} and \emph{Data Store} blocks (C9).

The absence of a category from one paradigm's literature is as analytically informative as its presence. C1 and C3, prominently referenced in traditional descriptions, are completely absent from AI-enabled literature. C10, C8, and C9, by contrast, are strongly present in AI descriptions. Ports, interfaces and subsystems (C6) is the exception, showing balanced modest reference in both ($9.7\%$ traditional, $10.7\%$ AI).

Traditional descriptions spread attention across multiple categories, while AI-enabled descriptions concentrate on the AI category and supporting infrastructures. Despite a comparable block count ($31$ versus $28$), the distribution differs. This indicates a shift from dynamics composition to algorithm and integration-centered structure. Core control dynamics (L1) is functionally central in both paradigms ($38.7\%$ traditional, $35.7\%$ AI). It is instantiated through C1 and C2 traditionally, and through C10 in AI literature. Traditional literature also emphasizes Signal Constraint Enforcement (L2, $16.1\%$) and Signal Preprocessing \& Computation (L4, $16.1\%$). AI literature instead emphasizes L4 ($28.6\%$) and Learning \& Adaptation (L8, $14.3\%$), a reallocation, not an omission of responsibility. L2 is entirely absent from AI-enabled descriptions. Traditional literature implements constraints as explicit blocks, such as saturation and rate limiters. AI-enabled literature instead handles them implicitly during training, via reward shaping, penalties, or environment design. They therefore never register as L2 in our extraction, which counts only explicit block references. Supervisory coordination (L5) and environment interaction (L9) are likewise more visible in AI descriptions ($7.1\%$ vs.\ $3.2\%$; $10.7\%$ vs.\ $6.5\%$). This reflects AI workflows' greater emphasis on subsystem structure.

Overall, traditional Simulink controller modeling is presented as a dynamics-driven composition, built primarily from continuous and discrete control primitives and explicit nonlinear or constraint elements. AI-enabled controller modeling, by contrast, is presented as an AI computation core surrounded by discrete-time integration, data handling, and abstraction scaffolding. It shows significantly less explicit structural emphasis on continuous-time dynamics and nonlinearities, and describes constraint enforcement less often as a distinct structural component.

\begin{tcolorbox}[boxsep=0pt,left=3pt,right=3pt,colback=white]
\textbf{RQ1:}
The literature implies a shift from structurally explicit architectures centered on dynamics composition and constraint blocks in traditional control to AI policy-centric computation with supporting interface structures, even though both paradigms share core control dynamics as a central functional role. By concentrating almost entirely on AI cores (C10) relative to its supporting categories, the literature raises questions on whether real-world control designs follow this narrow emphasis.
\end{tcolorbox}

\begin{table*}[!t]
\centering
\caption{Category-level structural comparison across controller paradigms. Controller subsystems across traditional families (PID, MPC, LQR, SBC, CLC), AI-enabled families (DNN, DRL, FLC), and aggregated Traditional (T) vs.\ AI.}
\label{tab:rq2_metric}
% \scriptsize
\fontsize{6.2}{10.5}\selectfont
\setlength{\tabcolsep}{2.0pt}
\renewcommand{\arraystretch}{0.8}
% \resizebox{\columnwidth}{!}{%
\begin{tabular}{c|rrrrrrrr|rr||rrrrrrrr|rr||rrrrrrrr|rr}
\toprule
& \multicolumn{10}{c||}{\textbf{Coverage ($\mathit{Cov_c}$, \%)}} & \multicolumn{10}{c||}{\textbf{Mean normalized presence ($\mathit{MNP_c}$, \%)}} & \multicolumn{10}{c}{\textbf{Model-level prevalence ($\mathit{MLP_c}$, \%)}} \\
\cmidrule(lr){2-11}\cmidrule(lr){12-21}\cmidrule(lr){22-31}

\textbf{Cat.} & \multicolumn{5}{c}{\textbf{Traditional}} & \multicolumn{3}{c|}{\textbf{AI-enabled}} & \multicolumn{2}{c||}{\textbf{All}} 
& \multicolumn{5}{c}{\textbf{Traditional}} & \multicolumn{3}{c|}{\textbf{AI-enabled}} & \multicolumn{2}{c||}{\textbf{All}} 
& \multicolumn{5}{c}{\textbf{Traditional}} & \multicolumn{3}{c|}{\textbf{AI-enabled}} & \multicolumn{2}{c}{\textbf{All}}\\
\cmidrule(lr){2-6}\cmidrule(lr){7-9}\cmidrule(lr){10-11}
\cmidrule(lr){12-16}\cmidrule(lr){17-19}\cmidrule(lr){20-21}
\cmidrule(lr){22-26}\cmidrule(lr){27-29}\cmidrule(lr){30-31}

& \textbf{PID} & \textbf{MPC} & \textbf{LQR} & \textbf{SBC} & \textbf{CLC} & \textbf{DNN} & \textbf{DRL} & \textbf{FLC} & \textbf{T} & \textbf{AI}
& \textbf{PID} & \textbf{MPC} & \textbf{LQR} & \textbf{SBC} & \textbf{CLC} & \textbf{DNN} & \textbf{DRL} & \textbf{FLC} & \textbf{T} & \textbf{AI}
& \textbf{PID} & \textbf{MPC} & \textbf{LQR} & \textbf{SBC} & \textbf{CLC} & \textbf{DNN} & \textbf{DRL} & \textbf{FLC} & \textbf{T} & \textbf{AI}\\
\midrule

C1 & 0.3 & 0.0 & 13.5& 0.0 & 0.0 & 0.7 & 0.0 & 0.6 & 0.4 & 0.3 
& 5.3 & 0.0 & 16.8& 0.0 & 0.0 & 1.7 & 0.0 & 0.4 & 3.9 & 0.2
& 45.5 & 0.0  & 75.0 & 0.0  & 0.0  & 33.3 & 0.0   & 20.0 & 25.8 & 6.5 \\

C2 & 0.7 & 0.7 & 0.0 & 3.1 & 0.3 & 0.3 & 5.5 & 0.6 & 0.7 & 3.5 
& 1.5 & 0.5 & 0.0 & 1.3 & 0.3 & 1.2 & 5.5 & 0.3 & 0.8 & 4.2 
& 63.6 & 33.3 & 0.0  & 33.3 & 28.6 & 33.3 & 100 & 20.0 & 38.7 & 80.6 \\

C3 & 0.6 & 0.0 & 5.8 & 0.8 & 0.0 & 0.2 & 0.0 & 1.2 & 0.5 & 0.2 
& 1.4 & 0.0 & 3.6 & 0.3 & 0.0 & 0.4 & 0.0 & 1.2 & 0.9 & 0.2 
& 72.7 & 0.0  & 25.0 & 33.3 & 0.0  & 66.7 & 0.0   & 40.0 & 32.3 & 12.9 \\

C4 & 1.1 & 0.0 & 0.0 & 4.6 & 12.5& 0.0 & 1.4 & 0.3 & 2.5 & 0.9 
& 1.5 & 0.0 & 0.0 & 1.9 & 10.5& 0.0 & 1.4 & 0.2 & 3.1 & 1.1 
& 36.4 & 0.0  & 0.0  & 33.3 & 100 & 0.0  & 100 & 20.0 & 38.7 & 77.4 \\

C5 & 7.6 & 5.7 & 4.8 & 6.9 & 14.8& 24.1& 1.4 & 7.1 & 8.1 & 8.4 
& 23.5& 5.4 & 9.4 & 2.9 & 14.2& 24.4& 1.4 & 5.9 & 14.0& 4.3 
& 100 & 100 & 50.0 & 33.3 & 57.1 & 100 & 100 & 80.0 & 77.4 & 96.8 \\

C6 & 72.3& 70.0& 52.9& 59.5& 51.3& 57.7& 80.7& 59.5& 68.8& 72.0 
& 52.3& 66.9& 42.8& 61.3& 48.1& 52.9& 80.7& 58.3& 54.6& 74.4 
& 100 & 100 & 75.0 & 100 & 100 & 100 & 100 & 100 & 96.8 & 100 \\

C7 & 6.8 & 5.0 & 5.8 & 8.4 & 9.9 & 12.3& 6.9 & 6.7 & 6.9 & 8.4 
& 5.4 & 5.6 & 5.4 & 3.6 & 13.2& 12.1& 6.9 & 7.5 & 6.9 & 7.5 
& 63.6 & 66.7 & 75.0 & 33.3 & 100 & 100 & 100 & 100 & 71.0 & 100 \\

C8 & 0.2 & 11.8& 4.8 & 5.3 & 0.0 & 0.4 & 2.8 & 8.6 & 2.0 & 2.7 
& 0.6 & 12.7& 4.2 & 13.0& 0.0 & 1.6 & 2.8 & 9.6 & 4.4 & 3.8 
& 36.4 & 100 & 25.0 & 100 & 0.0  & 33.3 & 100 & 100 & 45.2 & 93.5 \\

C9 & 10.5& 6.9 & 12.5& 11.5& 11.2& 4.3 & 1.4 & 15.3& 10.1& 3.7 
& 8.4 & 8.9 & 17.9& 15.6& 13.7& 5.7 & 1.4 & 16.5& 11.4& 4.2 
& 72.7 & 100 & 100 & 100 & 100 & 100 & 100 & 100 & 90.3 & 100 \\
\bottomrule
\end{tabular}%
%}
\end{table*}

\subsubsection{RQ2: Model-based composition}
\label{sec:rq2}
To evaluate RQ2, we analyze the structural composition of real-world controller subsystems through two complementary criteria: dominant categories and essential categories. Dominant categories are identified by coverage (Cov) and mean normalized presence (MNP). These metrics quantify how much of the controller structure is occupied by a category, across the dataset and within a typical controller, respectively. Essential categories are identified by model-level prevalence (MLP), which answers whether a category is present in almost all controllers regardless of its size contribution. Table~\ref{tab:rq2_metric} reports category-level measurements for controller subsystems extracted from $62$ Simulink CPS models spanning traditional and AI-enabled paradigms. C10 is not reported because AI controller cores are typically realized as nested or masked subsystems in Simulink. Their internal atomic blocks are counted under other categories and do not appear as standalone AI blocks at the controller level.

\textit{Dominant categories.} Coverage and MNP reveal a stronger pattern: subsystem organization, interfacing, and routing (C6) capture most controller structure (Cov: 68.8\% traditional, 72.0\% AI). C5 holds secondary structural mass by MNP in both paradigms (14.0\% traditional, 4.3\% AI); by Coverage, C9 is marginally higher than C5 in traditional models (10.1\% vs.\ 8.1\%). AI-enabled controllers allocate an even larger structural share to C6 than traditional controllers in MNP as well ($54.6\%$ traditional vs.\ $74.4\%$ AI). This indicates stronger centrality of hierarchical decomposition and interfacing in AI designs. Although C1 and C3 appear in some models, their dominance remains small in aggregate for both paradigms (C1: $0.4\%$/$0.3\%$; C3: $0.5\%$/$0.2\%$, traditional/AI). This suggests that explicit dynamics and nonlinearity blocks rarely consume substantial controller structure at the subsystem level.

To illustrate dominance patterns, we analyze family signatures. A family refers to a group of controller types within the same paradigm sharing a common design approach. For example, PID, MPC, and LQR form one family in the traditional paradigm. The family-level breakdown clarifies how dominance varies by controller type. Classical controllers concentrate their non-infrastructural structure in dynamics and computation categories. LQR assigns $13.5\%$ Cov to C1, while PID and MPC emphasize computation and infrastructure through C5 and C6 (C6 Cov: $72.3\%$ and $70.0\%$, respectively). Supervisory controllers expand logic and coordination, reflected in higher C4 in both SBC ($4.6\%$) and CLC ($12.5\%$). AI-enabled families shift structural dominance toward computation and architectural categories. DNN assigns substantial weight to C5 ($24.1\%$) and C6 ($57.7\%$). DRL assigns its dominant share to C6 ($80.7\%$). FLC distributes weight across C5, C6, and C9 (C9: $15.3\%$). DRL's family signature reflects a fixed Simulink scaffold applied identically across its 23 models regardless of plant, not 23 independently varied designs. This should be weighed when interpreting its outsized contribution to the AI-enabled aggregate.

\textit{Essential categories.} MLP results show that practical controller implementations consistently include categories that support architectural scaffolding and signal handling. C6 is nearly universal in both paradigms ($96.8\%$ traditional, $100.0\%$ AI). C5 is also widely adopted in both ($77.4\%$, $96.8\%$). Categories related to signal sources and stimuli (C7) and memory and data buffers (C9) show broad adoption as well ($71.0\%$/$90.3\%$ traditional, $100.0\%$ AI for both). In contrast, continuous dynamics (C1) and discontinuities and nonlinearities (C3) are not essential across the dataset. Both show lower MLP values in traditional and AI paradigms (C1: $25.8\%$/$6.5\%$; C3: $32.3\%$/$12.9\%$). This suggests selective use, not near-universal inclusion.

These patterns show that C5, C6, C7, and C9 form a shared structural core present in both paradigms. C1 and C10 are the primary paradigm-specific differentiators. This indicates that the shift from traditional to AI-enabled design is structurally localized, not pervasive.

\begin{tcolorbox}[boxsep=0pt,left=3pt,right=3pt,colback=white]
\textbf{RQ2:} C6 is near universal and structurally dominant across all controller families, reflecting heavy use of subsystems and interfacing in Simulink controller design. Beyond C6, C5, C7, and C9 are the most essential and dominant categories in both paradigms. AI-enabled controllers rely relatively more on C2, while traditional controllers rely more on C9, reflecting heavier explicit routing and buffering around control logic. This is a continuity that literature descriptions underrepresent but practitioners must navigate.
\end{tcolorbox}

\begin{figure}[t]
    \centering
    \includegraphics[width=0.5\textwidth]{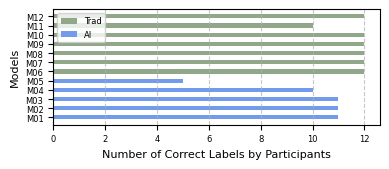}
    \caption{Survey results for controller design selection across benchmark models (n = 13 participants).} 
    \label{fig:survey}
\end{figure}

\begin{table}[!t]
\centering
\caption{Survey comparison of taxonomy categories: average presence (0--2) / average essence (1--5) per family (Traditional: PID, MPC, LQR, SBC, CLC; AI-enabled: DNN, DRL, FLC) and aggregated (T vs.\ AI).}
\label{tab:rq3_survey}
\scriptsize
\setlength{\tabcolsep}{1.8pt}
\renewcommand{\arraystretch}{1.5}
\begin{tabular}{c|ccccc|ccc|cc}
\toprule
& \multicolumn{5}{c|}{\textbf{Traditional}} & \multicolumn{3}{c|}{\textbf{AI-enabled}} & \multicolumn{2}{c}{\textbf{All}} \\
\cmidrule(lr){2-6}\cmidrule(lr){7-9}\cmidrule(lr){10-11}
\textbf{Cat.} & \textbf{PID} & \textbf{MPC} & \textbf{LQR} & \textbf{SBC} & \textbf{CLC} & \textbf{DNN} & \textbf{DRL} & \textbf{FLC} & \textbf{T} & \textbf{AI} \\
\midrule
C1 & 1.1/3.5 & 0.4/2.1 & 1.8/4.7 & 1.3/3.7 & 0.2/1.6 & 0.9/2.7 & 0.3/1.9 & 0.3/1.6 & 1.0/3.1 & 0.5/2.1 \\
C2,C4 & 1.0/3.3 & 1.0/3.5 & 0.3/2.2 & 1.5/3.9 & 1.2/3.9 & 0.9/2.9 & 1.9/4.8 & 0.9/3.1 & 1.0/3.3 & 1.2/3.6 \\
C3 & 0.9/3.1 & 0.4/2.3 & 0.3/1.9 & 1.2/3.5 & 0.2/1.9 & 0.7/2.8 & 1.0/3.2 & 1.0/3.3 & 0.6/2.5 & 0.9/3.1 \\
C8 & 0.5/2.3 & 1.8/4.9 & 1.3/4.0 & 0.6/2.6 & 0.9/3.3 & 1.1/3.3 & 1.6/4.5 & 1.6/4.6 & 1.0/3.4 & 1.4/4.1 \\
C7,C9 & 1.6/4.1 & 1.8/4.5 & 1.2/4.2 & 1.4/4.2 & 1.6/4.2 & 1.8/4.7 & 1.5/4.3 & 1.5/4.2 & 1.5/4.2 & 1.6/4.4 \\
\bottomrule
\end{tabular}
\footnotesize{\\
\textit{Cell format:} avg.\ presence / avg.\ essence.}
\end{table}

\subsubsection{RQ3: Practice-based corroboration}
\label{sec:rq3}

%To evaluate RQ3, we examined practitioners’ responses when interpreting controller designs extracted from the curated Simulink models. 
% Participants were asked to (i) identify whether each controller belonged to a traditional or AI-enabled paradigm and (ii) assess the structural influence of taxonomy categories when interpreting controller behavior. 
Survey responses are analyzed against the two tasks described in Section~\ref{sec:exp_setup}, using the five block groups from Table~\ref{tab:rq3_survey} and two aggregated metrics. \emph{Average presence} reflects how frequently participants selected a category on a scale of 0--2: not dominant $[0,0.9)$, moderately dominant $[0.9,1.5)$, or highly dominant $[1.5,2]$. \emph{Average essence} captures perceived category importance on a scale of 1--5: scarcely essential $[1,2.5)$, moderately essential $[2.5,3.5)$, or highly essential $[3.5,5]$. Fleiss' $\kappa$ across participants' paradigm classifications was $0.4847$ (moderate inter-rater agreement). Figure~\ref{fig:survey} and Table~\ref{tab:rq3_survey} summarize these classifications and scores. Higher values indicate stronger consensus that a category is structurally prominent and functionally central.

\textit{Manual Classification.} Most participants classified controller paradigms consistently: at least 11 of 13 identified traditional controllers, and 10--11 did so for most AI-enabled models. The exception is M05, a DNN-based controller correctly classified by only 5 of 13. This ambiguity stems from classical dynamics blocks coexisting with neural computation layers, and feedback-like signal flow resembling conventional architectures. It also reflects the absence of reward or memory loops expected of reinforcement learning. Similar ambiguity arose where traditional scaffolding surrounded AI components, showing how shared elements can obscure paradigm boundaries.

\textit{Dominant categories.} Table~\ref{tab:rq3_survey} shows C7,C9 as highly dominant in both paradigms (presence: $1.5$/$1.6$, traditional/AI). C8 and C2,C4 are moderately dominant, scoring higher for AI-enabled controllers (C8: $1.0$/$1.4$; C2,C4: $1.0$/$1.2$). C1 is moderately dominant in traditional but not AI-enabled controllers ($1.0$ vs.\ $0.5$). These match RQ2's model-based patterns. C2,C4's higher AI dominance links to C2's greater AI-family adoption, and C8's higher AI dominance links to its broader RQ2 adoption. Within families, C1 dominates LQR but not other traditional families. SBC and CLC show stronger C2,C4 dominance, reflecting recognition of discrete and logic-based elements as defining event-driven controllers. Among AI-enabled families, DRL associates most strongly with C2,C4 (matching C2's $100\%$ MLP in DRL models). DNN and FLC instead show stronger C8 dominance, reflecting user-defined abstraction blocks' role.

\textit{Essential categories.} Table~\ref{tab:rq3_survey} also shows C7,C9 rated as highly essential in both paradigms (essence: $4.2$/$4.4$, traditional/AI). C8 ($3.4$/$4.1$) and C2,C4 ($3.3$/$3.6$) are moderately to highly essential, with C8's elevated AI rating driven by families such as DRL and FLC. C1 is moderately essential in traditional but not AI-enabled controllers ($3.1$/$2.1$). C3 is scarcely to moderately essential in both ($2.5$/$3.1$). These largely align with RQ2. The high C7,C9 ratings match their near-universal MLP. The lower C1 rating in AI-enabled controllers matches its reduced MLP ($6.5\%$ AI vs.\ $25.8\%$ traditional). The higher C8 rating matches its higher MLP ($93.5\%$ AI vs.\ $45.2\%$ traditional). The main misalignment is C3, rated marginally more essential in AI-enabled than traditional controllers despite its lower MLP in AI families ($12.9\%$ vs.\ $32.3\%$). This likely occurs because practitioners interpret signal-shaping or transformation blocks as constraint mechanisms, conflating appearance with function.

\textit{Practice implications.} Practitioner judgment on C7 and C9 aligns closely with their measured model-level prevalence. This suggests practitioners can reliably judge essential structural elements, once controller structure is organized using the taxonomy's category groupings. The C3 misalignment, however, indicates a risk specific to AI-enabled controllers. Practitioners rate constraint-related structure as more essential than its actual prevalence supports, likely confusing signal-shaping or transformation blocks for constraint mechanisms. This risks overestimated constraint enforcement during manual safety review.

\begin{tcolorbox}[boxsep=0pt,left=3pt,right=3pt,colback=white]
\textbf{RQ3:} Manual practitioner assessment corroborates RQ1 and RQ2's structural patterns, though scaffolding-driven ambiguity persists. Dominant and essential categories align with RQ2: C7,C9 are broadly indispensable, and C1 is correctly less central for AI-enabled designs. The main misalignment is C3, rated more essential than its low model-level prevalence supports.
\end{tcolorbox}

\subsection{Discussion}

Our empirical evaluation revealed three tensions: 
%(1) between what literature emphasizes and what models actually contain, (2) between the continuity of structural scaffolding across paradigms and the perception of disruption, and (3) between the invisible treatment of constraints in AI-enabled controllers and practitioner expectations. We examine each tension as a lesson for MBSE below.

\textbf{Lesson 1: The ``Scaffolding'' Dominates the System Model}

The most striking finding is the sheer dominance of what we term architectural ``scaffolding'' across all controller paradigms. C6, the primary scaffolding category, consistently dominates all controller families regardless of paradigm, occupying the largest share of controller structure. C5 plays a secondary role, broadly present but varying in weight. This reflects a fundamental MBD truth: control systems embody an architecture, not just an algorithm, requiring deep design knowledge to realize. In contrast, the control law, whether C1-based or an AI policy based on C10, occupies a small fraction of the model's structure. Most of it is dedicated to signal routing, subsystem hierarchy, data buffering, and interfacing. This integrates the control component into the system's broader environment.

Scaffolding's composition also shifts between paradigms. AI-enabled families most consistently adopt categories managing sampling, signal movement, and custom computation (C2, C7, C8), introducing distributed dependencies around the learned component. Traditional designs instead cluster non-infrastructure structure more narrowly, C1 in LQR, C4 in supervisory controllers, C5 in PID and MPC, yielding more concentrated family signatures. Scaffolding remains dominant, but its composition evolves. This challenges the MBSE community. Migrating from a traditional to an AI-enabled controller raises not just which blocks to swap, but how much scaffolding must be re-engineered. This re-engineering effort falls on the majority share of the model, not the control algorithm itself. Tooling that treats scaffolding as a reusable, first-class abstraction, instead of ad-hoc blocks, could directly reduce the cost of paradigm migration. This would let engineers focus effort on the smaller, paradigm-specific portion of the design.

\textbf{Lesson 2: The Shift is Less Radical in Practice}

A key tension emerges between AI-focused literature and real-world models. The literature presents AI-enabled control as a major architectural shift centered on AI-specific components and their infrastructure. Real-world models instead show much stronger structural continuity: traditional and AI-enabled controllers share a common architectural core. Differences often reduce to a few paradigm-specific elements that make hybrid models especially difficult for practitioners to interpret consistently. RQ2 shows C5, C6, C7, and C9 forming this common core. RQ3 confirms practitioners perceive C7 and C9 as highly dominant in both. The primary differentiator is the presence or absence of C1 and C10, embedded within an otherwise familiar context.

We further examined M05, misclassified by 8 participants (Figure~\ref{fig:survey}). This case reflects incremental AI integration in MBSE. Its AI component was embedded within traditional scaffolding, including classical dynamics blocks and feedback loops resembling traditional architectures. This ambiguity does not indicate a failure in practitioner perception. It instead reflects a common industrial migration pattern, where engineers integrate new AI components into existing structures. Shared scaffolding is what makes hybrid designs hard to classify. Modeling patterns that visually or structurally distinguish learned components from surrounding scaffolding could directly reduce this ambiguity for practitioners performing manual review.

\textbf{Lesson 3: Mind the Architectural Blind Spots}

For the safety and verification community, constraint enforcement is treated very differently across paradigms. Traditional designs encode constraints as explicit structural elements. AI-enabled systems instead handle them during training, leaving them invisible in the deployable architecture and creating a serious traceability gap. Practitioners are not well positioned to catch this gap through manual inspection alone. RQ3 shows they rate constraint-related structure as more essential than its actual prevalence supports, likely mistaking signal-shaping blocks for constraint mechanisms. Safety-relevant logic has moved to training-time mechanisms invisible in the model, and practitioners cannot reliably detect this absence through inspection. Verification processes relying on manual review of model structure are likely to miss this gap.

These three tensions point to two concrete needs for MBSE tool support. First, scaffolding dominates controller structure and must be substantially re-engineered during paradigm migration; tools should represent it as a reusable, first-class abstraction instead of boilerplate. Second, safety-relevant constraint logic in AI-enabled controllers moves to training-time mechanisms invisible during manual review. Tools should represent reward-shaping, penalty, and environment-design decisions as explicit, inspectable artifacts. This mirrors how traditional designs already represent constraints as explicit blocks. We leave designing and validating such tooling to future work.

%% file: Sections/Threats.tex
\textit{Construct Validity.} A major threat is whether our taxonomy and metrics capture controller architecture or merely general Simulink conventions. C6's dominance in both paradigms may reflect modeling style more than design differences. We mitigate this by basing the taxonomy on literature, validating it against Simulink's library structure, and triangulating across metrics and human interpretation. A second threat, treating literature block counts as a proxy for structural composition, is addressed by cross-checking literature-based against model-based findings for mismatches. A further threat concerns the AI-enabled sample: 23 of 31 models are DRL controllers on a standardized benchmark wrapper. This yields near-identical composition across plants, with multi-agent variants scaling it by an integer factor. Coverage, aggregating at block level, is thus more sensitive to this than $\mathit{MNP_c}$ and $\mathit{MLP_c}$, which weight models equally regardless of size. This homogeneity likely reflects current DRL tooling, where controllers are typically built from standardized RL wrappers rather than bespoke designs, not dataset-specific. We report DRL separately from DNN and FLC in Table~\ref{tab:rq2_metric} to keep its homogeneity visible instead of folded into one aggregate.

\textit{Internal Validity.} Literature review may omit relevant sources; we included peer-reviewed publications and grey literature, cross-validated across multiple sources. The model dataset may contain incomplete subsystems; we applied exhaustive block extraction across nested subsystems, masked components, and linked libraries. We also used multiple researchers in taxonomy development, resolving disagreements through discussion. Taxonomy development was collaborative, not independently coded, so we did not compute a formal inter-rater agreement measure such as Cohen's $\kappa$. Consistency was instead ensured through discussion-based consensus among researchers. Participant expertise varies; we collected self-reported experience data and required minimum control-systems knowledge. Using screenshots instead of interactive models may limit what participants infer about system structure, mitigated with consistent annotations and author contact for clarification.

\textit{External Validity.} Our 62 Simulink models, drawn from open-source repositories, benchmarks, and competition datasets across several CPS domains, may not fully represent proprietary industrial practice. We mitigate this through diverse controller families and paradigm balance. Findings may not generalize beyond Simulink or to other domains (e.g., power systems). The 13-participant sample may also not capture the full range of industrial expertise, bounded by reporting demographic characteristics.

\textit{Conclusion Validity.} Our conclusions are limited by the dataset, literature, and survey participants. Explicit block references may not reflect full controller structure, and participant judgments may vary with background and experience. We reduce these risks by combining literature, model, and human-centered evidence and designing the study for consistent comparison across controller types.

%% file: Sections/Related.tex
\textit{AI for Model-based Engineering.}
Several studies examine how AI can enhance and automate model-based practices. These range from AI-enhanced modeling feasibility~\cite{Song2025AI4MBSE} to frameworks combining knowledge graphs and retrieval-augmented generation (RAG) for AI-augmented engineering~\cite{Hanke2025AIAugmentedMBSE}. Work in Model-Driven Engineering (MDE) has separately reviewed domain-specific languages and practices for AI software systems~\cite{Radler2024MDEAI}. This work highlights limited MDE-AI-DevOps integration in CPS, with support confined to isolated development stages, a gap our work addresses through architectural characterization.

\textit{Empirical and Taxonomy-Based Studies of Simulink Models.} Boll et al.~\cite{Boll2021OpenSourceSimulink} show that model size and block counts are insufficient to infer architectural intent or control design choices. Shrestha et al.~\cite{Shrestha2023ReplicabilitySimulinkCorpora} highlight the difficulty of replicating Simulink-based studies, motivating carefully curated corpora. Taxonomy studies in Simulink remain limited, largely classifying specific engineering concerns instead of controller architecture. Elberzhager et al.~\cite{Elberzhager2013SimulinkMapping} organize testing techniques, and Weiland et al.~\cite{Weiland2014SimulinkVariabilityClassification} classify variability mechanisms. None characterize how controller designs are structurally realized, nor compare traditional and AI-enabled architectures.

\textit{AI vs. Traditional Control in Simulink-Based CPS.}
A growing body of work compares AI-based and traditional controllers through Simulink implementations. Song et al.~\cite{song2022cyber} and Okafor et al.~\cite{Okafor2021RLBasedPIDvsClassical} evaluate learning-based controllers against classical designs such as PID, often reporting adaptability or performance gains. Song et al. also provide a benchmark suite now widely used for comparative AI-enabled CPS studies. This work centers on control performance, treating structure as secondary. It does not analyze how control responsibilities are distributed across Simulink blocks and subsystems, or how AI-enabled controllers structurally differ from traditional ones. As Yusuf and Gaaloul~\cite{hadiza2025Nav} note, these architectural shifts and their verification implications remain undercharacterized. Related debugging techniques for CPS behavior, such as counterfactual-guided assertion inference~\cite{ghazal2026decaf}, similarly depend on understanding which model components drive observed failures. Verification approaches for AI-enabled CPS, such as guided falsification of temporal logic properties~\cite{yusuf2025guidedfalsification}, likewise depend on understanding the underlying controller structure.

%% file: Sections/Conclusion.tex
We presented an empirical study of Simulink-based CPS controllers across traditional and AI-enabled paradigms. This combined a literature-derived taxonomy, analysis of 62 real-world models, and a practitioner survey. Results show that subsystem organization dominates all controller structures. AI-enabled controllers rely heavily on discrete dynamics and user-defined abstraction, despite underspecification in the literature. Constraint enforcement blocks are structurally absent from AI-enabled models, despite practitioner expectations. These findings establish the first empirical baseline for cross-paradigm architecture comparison, motivating paradigm-aware tool support for verification and safety assurance in AI-enabled CPS.

% \section{Data Availability Statement}
% All data and artifacts are available in our replication package~\cite{anonymous_zenodo_2026}.